\documentclass[a4paper]{article}
\usepackage{multirow}
\usepackage{booktabs}
\usepackage{ctable}
\usepackage{INTERSPEECH2020}

\title{Towards Natural Bilingual and Code-Switched Speech Synthesis Based on Mix of Monolingual Recordings and Cross-Lingual Voice Conversion}
\name{Shengkui Zhao, Trung Hieu Nguyen, Hao Wang, Bin Ma}
\address{Machine Intelligence Technology, Alibaba Group}
\email{\{shengkui.zhao, trunghieu.nguyen, hao.w, b.ma\}@alibaba-inc.com}

\begin{document}

\maketitle
\begin{abstract}
Recent state-of-the-art neural text-to-speech (TTS) synthesis models have dramatically improved intelligibility and naturalness of generated speech from text. However, building a good bilingual or code-switched TTS for a particular voice is still a challenge. The main reason is that it is not easy to obtain a bilingual corpus from a speaker who achieves native-level fluency in both languages. In this paper, we explore the use of Mandarin speech recordings from a Mandarin speaker, and English speech recordings from another English speaker to build high-quality bilingual and code-switched TTS for both speakers. A Tacotron2-based cross-lingual voice conversion system is employed to generate the Mandarin speaker's English speech and the English speaker's Mandarin speech, which show good naturalness and speaker similarity. The obtained bilingual data are then augmented with code-switched utterances synthesized using a Transformer model. With these data, three neural TTS models -- Tacotron2, Transformer and FastSpeech are applied for building bilingual and code-switched TTS. Subjective evaluation results show that all the three systems can produce (near-)native-level speech in both languages for each of the speaker.
%Recent state-of-the-art neural text-to-speech (TTS) synthesis models have dramatically improved intelligibility and naturalness of generated speech from text. However, building a good bilingual or code-switched TTS (in a single speaker's voice) is still a challenge. One reason is that it is not easy to obtain a bilingual corpus from a speaker who achieves native-level fluency in both languages. In this paper, we explore the use of two mononlingual speech recordings from a Mandarin speaker and an English speaker only, to build high-quality bilingual and code-switched TTS for both speakers. A Tacotron2-based cross-lingual voice conversion system is employed to generate the Mandarin speaker's English speech and the English speaker's Mandarin speech, which show good naturalness and speaker similarity. Then, the obtained bilingual data are further augmented with code-switched speech synthesized using a Transformer model. With these data, three neural TTS models -- Tacotron2, Transformer and FastSpeech are applied for building bilingual and code-switched TTS. Subjective evaluation results show that all the three systems can produce (near-)native-level speech in both languages for each of the speaker.
\end{abstract}
\noindent\textbf{Index Terms}: cross-lingual voice conversion, Phonetic PosteriorGrams (PPGs), Tacotron2, Transformer, FastSpeech, text-to-speech, bilingual, code-switching

\section{Introduction}

State-of-the-art end-to-end neural TTS models \cite{Jonathan2018, Naihan2018, Ren2019} have been developed to produce high intelligible and natural monolingual speech in a single speaker’s voice. However, it is still a major challenge to extend such models to support bilingual and code-switched TTS, which is observed as growing demands in various scenarios. The difficulty in obtaining a bilingual corpus produced by a speaker who is highly proficient in both languages makes the task not straightforward. Therefore, quite a few efforts attempt to leverage on speech corpora from two monolingual speakers in different languages for building bilingual or code-switched TTS systems.

Early studies on TTS supporting more than one language with a monolingual speaker’s voice are mostly HMM-based methods. \cite{Latorre2005} proposes a Polyglot synthesis method that adapts the shared HMM states trained on a mixture of monolingual corpora to the target speaker. \cite{zen2012} proposes an HMM-based parametric TTS system based on a speaker and language factorization. \cite{JHe2012} uses a trajectory tiling approach to render the target monolingual speaker's speech waveforms in a second language. A HMM-based TTS is then built with the rendered speech and the target speaker's original recordings in the other language. \cite{Ramani2014} proposes using a GMM-based cross-lingual voice conversion (VC) to generate a monolingual speaker's speech in other languages. Eventually, this speaker's multilingual data is created, and then used for building an HMM-based polyglot speech synthesizer. \cite{Sitaram2016} presents an HMM-based parametric code-switched TTS system based on monolingual datasets. A combined phonetic space in both languages is explored, and the pronunciations are mapped across languages. \cite{FLXie2016} uses a speaker-independent DNN ASR output to map the senones between two monolingual corpora in two languages for building an HMM-based TTS system.

Recently, neural network based TTS models are also explored for cross-lingual TTS. \cite{Nachmani2019} presents a multi-lingual and multi-speaker neural TTS model based on the VoiceLoop architecture \cite{Taigman2018} with speaker and language embedding networks. \cite{YWCao2019} proposes code-switched TTS using Tacotron-based end-to-end systems \cite{YWang2017}. It uses a Mandarin and an English monolingual speech corpora uttered by two female speakers. Two mechanisms are implemented: (1) a shared encoder with language embedding, and (2) two separate language-dependent encoders. Experiments show that their systems work well when synthesizing an American speaker’s Mandarin speech, while the performance is not too good when synthesizing a Mandarin speaker's English speech. \cite{LXue2019} explores a Mandarin/English code-switched TTS model based on the Tacotron2 model \cite{Jonathan2018}. The authors investigate speaker embedding and phoneme-informed attention. They build a base model on multi-speaker monolingual data, and then adapt it to a target Mandarin speaker's voice. Their model is able to produce code-switched speech, but unnatural prosody and inaccurate tones can be observed for Mandarin words next to English words. Most recently, \cite{YZhang2019} presents a multi-lingual TTS model that is able to preserve a monolingual target speaker's voice characteristics across three different languages based on Tacotron2 \cite{Jonathan2018}. The model is trained on monolingual recordings from a large number of speakers. It uses a unified phoneme input representations, and also incorporates an adversarial loss to decouple speaker identities from speech content. However, their model relies on data from a large number of speakers per language to achieve good performance in cross-lingual voice cloning; and it only provides rudimentary support for code-switching.

In this work, we present a method to build bilingual and code-switched TTS models using monolingual corpora from a female British English speaker and a female Mandarin speaker. We obtain the Mandarin speaker’s English speech and the English speaker’s Mandarin speech by a Tacotron2-based cross-lingual voice conversion method. Although the idea of using cross-lingual VC for mixed-lingual TTS has been presented in \cite{Ramani2014}, our cross-lingual VC method achieves good naturalness and speaker similarity. Using each speaker’s original corpus and the converted speech in the other language, we explore different neural TTS model architectures -- Tacotron2 \cite{Jonathan2018}, Transformer \cite{Naihan2018} and FastSpeech \cite{Ren2019} for bilingual and code-switched speech synthesis. By using Transformer TTS  models trained on the obtained bilingual corpora to generate code-switched speech to further augment the training data, our experiments and evaluation results show that all the three systems can produce (near-)native-level speech in both languages and on code-switched content as well. Our contributions include: (1) to the best of our knowledge, this is the first work that adapts neural TTS model architectures to cross-lingual VC in building bilingual and code-switching TTS, where high speech quality and speaker similarity are achieved; (2) we explore and implement three different models for bilingual and code-switched TTS.

\begin{figure}[t]
  \centering
  \includegraphics[width=\linewidth]{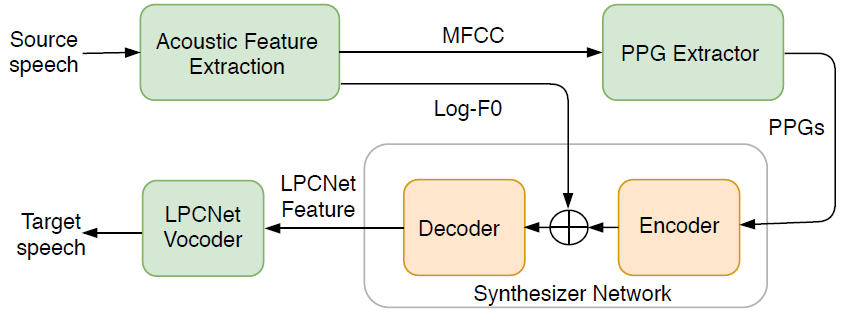}
  \caption{Our neural framework for cross-lingual voice conversion of source speech to target speech.}
  \label{fig1}
\end{figure}

\section{Building Bilingual and Code-Switched TTS}

This work uses monolingual corpora from an English speaker and a Mandarin speaker. The objective is to build high-quality English-Mandarin bilingual and code-switched TTS for each of the speaker's voice. One of the keys to building such a TTS with monolingual data is to solve bilingual phonetic coverage \cite{LXue2019}. We realize a full bilingual phonetic coverage by cross-lingual VC for both speakers, that is, generating the Mandarin speaker's English speech and the English speaker's Mandarin speech. Thus, each of the original monolingual corpora is expanded to be bilingual. Based on these data, we then explore three architectures -- Tacotron2 \cite{Jonathan2018}, Transformer \cite{Naihan2018} and FastSpeech \cite{Ren2019} for bilingual and code-switched TTS.

\subsection{Building bilingual corpora with Tacotron2-VC}
The speech data that we have are two monolingual corpora from a Mandarin speaker and an English speaker. Cross-lingual voice conversion is employed to transform the voice of the original Mandarin corpus to the English speaker's voice and transform the voice of the original English corpus to the Mandarin speaker's voice. Thus both speakers' bilingual speech data are obtained. Different from other cross-lingual VC methods \cite{Mashimo2002, qian2011, Erro2010, wang2015, sun2016, zhou2019, 2019skzhao}, we adapt state-of-the-art TTS model -- Tacotron2 \cite{Jonathan2018} to this cross-lingual VC task. Together with LPCNet neural vocoder, this Tacotron2-VC framework is able to generate a target monolingual speaker's high-quality speech in another language. The whole framework is illustrated in Fig. 1. First, we use a Phonetic PosteriorGram (PPG) extractor to extract PPGs from a source speaker's utterances. We then generate the LPCNet features by feeding the PPGs to a Tacotron2-based synthesizer network that is well trained for mapping from PPGs to LPCNet features in a target speaker's voice. In this synthesizer network for VC, normalized log-F0 features from the source utterances are concatenated with the encoder output, for preserving the shape of F0 contours during the conversion. The detailed structure of Tacotron2-based synthesizer network for VC is shown in Fig. 2. Finally, a LPCNet vocoder that is trained on the target speaker's data convert the Tacotron2-VC-generated LPCNet features to waveforms. The components of Tacotron2-VC are elaborated in the following subsections.
\begin{figure}[t]
  \centering
  \includegraphics[width=5.5cm]{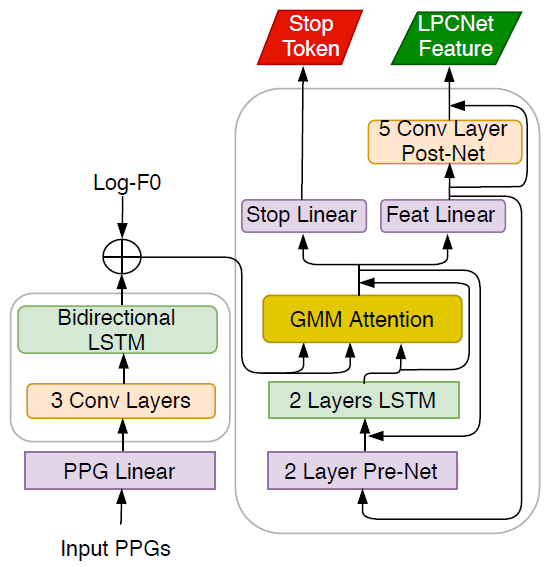}
  \caption{The architecture of the synthesizer network in Tacotron2-VC.}
  \label{fig2}
\end{figure}
\subsubsection{The PPG extractor}
A PPG represents posterior probabilities of phonetic classes (phonemes or triphones/senones) for every frame of an utterance \cite{hazen2009, kintzley2011}. It can be considered as a speaker- and language-independent representation of acoustic information, and is first used for cross-lingual VC in \cite{sun2016} and followed by \cite{zhou2019}. Limited by the converter models and vocoders used in their efforts, the quality of the generated speech is far lower than professional speech recordings. In Tacotron2-VC, we also use PPGs to bridge across the speaker and language differences. They are extracted by a PPG extractor. It works similarly as a speaker-independent acoustic model of an ASR system. The PPG extractor is trained to classify frame-based MFCCs to the corresponding senones by minimizing the cross-entropy loss. This PPG extractor is implemented based on the open-source package, Pytorch-Kaldi \cite{Ravanelli2019}. We train a network that contains 5 bidirectional GRU layers with 550 hidden units on each layer. The senone labels are obtained from force-alignments of a well-trained GMM-HMM system. In our experiments, we use 488-dim PPGs.

\subsubsection{The Tacotron2-based synthesizer network for VC}
As shown in Fig. 2, the synthesizer network in Tacotron2-VC is used for a sequence to sequence prediction from PPGs to LPCNet features. The architecture of the synthesizer network is composed of an encoder and a decoder with an attention mechanism. An input PPG is first processed by a fully connected feed-forward (FF) linear layer (PPG Linear). The PPG Linear has 512 output units, followed by the layer normalization and ReLU activation. Thus the PPG input is transformed to 512-dim embeddings. The output of the PPG linear is then converted into a hidden feature by the encoder, which is composed of a 3-layer 1D CNN ($512$ filters with shape $5 \times 1$) and 1-layer bidirectional LSTM ($256$ units in each direction). The hidden feature representation is concatenated with a 1-dim Log-F0 sequence and then passed to the decoder to predict 20-dim LPCNet features and 1-dim stop token. The decoder is composed of a Pre-Net (2 FF layers of 256 hidden ReLU units), a 2-layer LSTM of $1024$ units, 2 separate linear layers, and a 5-layer CNN ($512$ filters with shape $5 \times 1$) with a residual connection. We perform downsampling for the input PPG to reduce its length. We also replace the location-sensitive attention by the GMM attention \cite{Alex2013} to improve the alignment performance. Our Tacotron2-VC is implemented based on the open-source repository\footnote{https://github.com/keithito/tacotron}.

\subsubsection{The LPCNet vocoder}
A sequence of LPCNet feature vectors generated by the synthesizer network is reconstructed to a waveform by a LPCNet vocoder \cite{Valin2019}. Each LPCNet feature vector consists of 18 Bark-scale cepstral coefficients and 2 pitch parameters (period, correlation). We use the open-source code published by Mozilla team on Github  \footnote{https://github.com/mozilla/LPCNet} for the LPCNet vocoder.

\subsubsection{Training and conversion}
Our Tacotron2-VC consists of a training stage and an inference stage. In the training stage, all the above three components are trained separately.  The PPG extractor is speaker-independent and is used for PPG extraction on both the English and the Mandarin corpora. It is trained using AISHELL corpus \cite{aishell2017}. The synthesizer network for VC and the LPCNet vocoder are speaker dependent and trained on each speaker's original corpus. For each speaker's corpus, we extract the utterance-based PPGs, Log-F0s, and LPCNet features. The Log-F0s are normalized per speaker with zero mean and unit variance. The PPGs, normalized Log-F0s, and LPCNet features are used to train a synthesizer network for each speaker. The LPCNet features and the waveforms of the original speech recordings are used to train speaker-dependent LPCNet vocoders. When all the three components are well trained, we use the system to perform voice conversion on each corpus. For obtaining the Mandarin speech in the English speaker's voice, the Mandarin recordings are first processed to obtain MFCCs and Log-F0s, the MFCCs are sent to the PPG extractor to obtain PPGs. Both PPGs and normalized Log-F0s are fed into the converter trained on English recordings, to generate LPCNet features. Finally, these LPCNet features are passed to the LPCNet vocoder of the English speaker to generate Mandarin speech in the English speaker's voice. The same procedure applies for obtaining the English speech in the Mandarin speaker's voice.

\subsection{Bilingual and code-switched TTS models}
Next step is to build bilingual and code-switched TTS using the bilingual corpora obtained from the above conversion step. For each speaker, we apply three different model architectures including Tacotron2, Transformer and FastSpeech. Note that there is no code-switched utterances in the obtained bilingual corpora. The TTS models still need to learn the code-switching from monolingual English and Mandarin utterances. We describe our training procedure as follows.
\subsubsection{Input representation}
The typical input representations for end-to-end TTS models are character, phoneme, or UTF-8 byte encoding. In this work, we use phoneme representation as the work \cite{YZhang2019} suggests that the phoneme-based TTS model performs significantly better than char- or byte-based variants due to rare or OOV words. Instead of using a unified phone set across languages, we combine English and Mandarin phone sets together as a whole. For English utterances, we use 44 British English phoneme symbols plus 3 possible stress symbols. For Mandarin utterances, we use 62 Pinyin initials and finals plus 5 possible tones. The tone or stress symbols are attached to the corresponding phoneme symbols. We also use symbols to indicate in-utterance pauses and utterance ends.
\subsubsection{Training TTS models}
We use a Tacotron2 architecture based on the description in \cite{Jonathan2018}. But we replace the character input sequence by the phoneme input sequence. The original model predicts 80-dim mel-scale spectrograms and uses a WaveNet vocoder to synthesize time-domain waveforms from those spectrograms. We modify the model to predict 20-dim LPCNet features and use a LPCNet vocoder for waveform generation. Our implementation is based on the open-source code\footnote{https://github.com/keithito/tacotron}. For the Transformer TTS model, we use the architecture described in \cite{Naihan2018}. The original model also predicts mel-scale spectrograms for a WaveNet vocoder. Again, we modify the model to predict LPCNet features for a LPCNet vocoder. We also follow \cite{Ren2019} for the FastSpeech TTS model architecture. Similarly, we modify the model to generate LPCNet features. We use the open source code \emph{ESPnet}\footnote{https://github.com/espnet/espnet} to train the Transformer and FastSpeech models. As the FastSpeech model needs the phone duration as a target to train the duration predictor, we first train a Transformer TTS model to obtain encoder-decoder attention alignments on the training data sets. The obtained alignments are then used to train the duration predictor of the FastSpeech TTS model. In our experiments, we observe that the Tacotron2 TTS model trained only on the bilingual corpus produces more prosodic errors for code-switched text than the Transformer and the FastSpeech TTS models. We therefore use the Transformer TTS model to create a set of code-switched speech data for each of the two speakers. We then add these code-switched utterances into the bilingual training sets to refine the Tacotron2 TTS model. We find that such data augmentation process can also benefit the Transformer and FastSpeech TTS models.

\section{Experiments}

\subsection{Experimental setup}
Both the English and the Mandarin corpora in our experiments are professional speech recordings for TTS. The English corpus is produced by a female native British English speaker. It has 27,000 utterances and the total length is about 41 hours. The Mandarin corpus is produced by a female native Mandarin speaker. It has 32,000 utterances and the total length is about 30 hours. We select 250 utterances for validation and 250 utterances for testing. All speech data are sampled at 16 kHz with 16-bit resolution. We also create a code-switched text corpus of 17,000 sentences by replacing the selected English or Chinese words in monolingual sentences by their translated counterparts. We then use our well-trained Transformer TTS model to generate speech from these code-switched texts. The total length of obtained code-switched speech data set is about 16 hours. These generated code-switched utterances are used to refine the Tacotron2 TTS model.
\begin{table}[]
\caption{The cross-lingual VC naturalness and similarity MOS with 95\% confidence intervals for both English and Mandarin speakers using Tacotron2-VC (GT: Ground Truth). }
\begin{tabular}{lcccc}
\specialrule{.1em}{.05em}{.05em}
\multirow{3}{*}{\begin{tabular}[c]{@{}l@{}}Target \\ Speaker\end{tabular}} & \multicolumn{2}{c}{\multirow{2}{*}{MOS Scores}} \\
                                                                           & \multicolumn{2}{c}{}                           \\ \cline{2-3}
                                                                           & Naturalness (GT)            & Similarity (GT)            \\ \hline
Mandarin                                                                   & 3.99±0.07 (4.45±0.15)  & 3.79±0.11 (4.38±0.09) \\ \hline
English                                                                    & 4.11±0.10 (4.22±0.13)  & 3.71±0.09 (4.25±0.09) \\
\specialrule{.1em}{.05em}{.05em}
\end{tabular}
\end{table}
\subsection{Subjective tests}
We conduct MOS evaluations for both VC and TTS results. The VC results are evaluated on speech naturalness and speaker similarity separately, while the TTS results are evaluated on overall speech quality. Ratings follow the Absolute Category Rating scale, with scores from 1 (bad) to 5 (excellent) in 0.5 point increments. For speech naturalness tests, all raters are native speakers in the language of the evaluating utterances. For speaker similarity tests, most of the raters are Mandarin-English bilingual speakers. Rating on speaker similarity across languages is rather challenging. The raters are advised to ignore the speech content but focus only on the speaker identity. The ground truth are self similarity scores on the original speech recordings.
\subsubsection{VC results}
Two Tacotron2-VC system are trained separately for the two speakers. We convert the Mandarin recordings to the English speaker's voice and the English recordings to the Mandarin speaker's voice. Table 1 gives the naturalness and similarity MOS results for the converted speech in the two speakers voices. Each MOS score is averaged over 20 randomly picked utterances where each utterance is rated by 10 raters. The MOS scores in Table 1 indicate that the converted speech achieves reasonably good naturalness and similarity compared to the ground-truth MOS scores. Our following evaluation results on TTS also confirm the effectiveness of using the speech data converted from Tacotron2-VC for bilingual and code-switched speech synthesis.
\subsubsection{TTS results}
% \usepackage{multirow}
\begin{table}[]
\caption{The MOS with 95\% confidence intervals on the Mandarin speaker's bilingual and code-switching neural TTS models. }
% \usepackage{multirow}
\begin{tabular}{lccc}
\specialrule{.1em}{.05em}{.05em}
\multirow{2}{*}{Model} & \multicolumn{3}{c}{Language}        \\ \cline{2-4}
                       & Chinese   & English   & Code-Switch \\ \hline
GroundTruth            & 4.16±0.10                   & 4.20±0.15                   & -                               \\ \hline
Tacotron2              & 4.12±0.06                   & 3.95±0.11                   & 3.96±0.08                       \\ \hline
Transformer            & 4.03±0.06                   & 3.87±0.09                   & 3.97±0.08                       \\ \hline
FastSpeech             & 4.10±0.06                   & 3.91±0.08                   & 3.94±0.08                       \\
\specialrule{.1em}{.05em}{.05em}
\end{tabular}
\end{table}

\begin{table}
\caption{The MOS with 95\% confidence intervals on the English target speaker's bilingual and code-switching neural TTS models. }
\begin{tabular}{lccc}
\specialrule{.1em}{.05em}{.05em}
\multirow{2}{*}{Model} & \multicolumn{3}{c}{Language}        \\ \cline{2-4}
                       & Chinese   & English   & Code-Switch \\ \hline
GroundTruth            & 4.15±0.11 & 4.34±0.13 & -           \\ \hline
Tacotron2              & 4.07±0.06 & 4.09±0.09 & 3.99±0.08   \\ \hline
Transformer            & 4.12±0.07 & 4.02±0.09 & 3.98±0.07   \\ \hline
FastSpeech             & 4.08±0.07 & 4.16±0.10 & 3.90±0.08   \\
\specialrule{.1em}{.05em}{.05em}
\end{tabular}
\end{table}

Table 2 and 3 show the MOS results on speech quality of the synthesized utterances from the Tacotron2, Transformer, and FastSpeech TTS models. For monolingual English and Chinese utterances, raters are advised to focus on speech intelligibility and naturalness. For code-switched utterances, raters are advised to also focus on speaker consistency. From the results, we observe the synthesized monolingual Mandarin and code-switched speech in the English speaker's voice has comparable performance to that in the Mandarin speaker's voices. The MOS scores of the synthesized English speech in the English speaker voice are slightly higher than those on the synthesized English speech in the Mandarin speaker's voice. For the Mandarin speaker, the MOS scores of the synthesized Mandarin speech are slightly higher than those of the synthesized English and the code-switched speech. For the English speaker, the MOS scores of the synthesized English speech are similar as those of the synthesized Mandarin speech, and slightly higher than those of the synthesized code-switched speech. For both speakers, all the TTS MOS scores are close to the ground-truth MOS scores. The results indicate that the TTS models achieve good performance in synthesizing bilingual and code-switched speech. The readers are recommended to listen to our audio samples from the demo page\footnote{Audio samples - https://alibabasglab.github.io/tts/}.

\section{Conclusions}

In this paper, we present an approach to natural bilingual and code-switched TTS. This work relies only on two monolingual corpora. A Tacotron2-based cross-lingual VC is used to generate high-quality speech in the other language, thus expanding the two monolingual corpora to be bilingual ones. We show that the Transformer and FastSpeech TTS models that are trained on the generated bilingual corpora can synthesize natural bilingual and code-switching speech, of which the overall quality is close to the professional speech recordings. The bilingual corpora can be further augmented with the code-switched speech synthesized using Transformer model; then, these augmented corpora can be used to refine the Tacotron2 TTS model for achieving a comparable performance to the Transformer and FastSpeech models. The proposed framework of bilingual and code-switched TTS should be applicable to other speaker and language pairs.

\bibliographystyle{IEEEtran}

\bibliography{mybib}

% \begin{thebibliography}{9}
% \bibitem[1]{Davis80-COP}
%   S.\ B.\ Davis and P.\ Mermelstein,
%   ``Comparison of parametric representation for monosyllabic word recognition in continuously spoken sentences,''
%   \textit{IEEE Transactions on Acoustics, Speech and Signal Processing}, vol.~28, no.~4, pp.~357--366, 1980.
% \bibitem[2]{Rabiner89-ATO}
%   L.\ R.\ Rabiner,
%   ``A tutorial on hidden Markov models and selected applications in speech recognition,''
%   \textit{Proceedings of the IEEE}, vol.~77, no.~2, pp.~257-286, 1989.
% \bibitem[3]{Hastie09-TEO}
%   T.\ Hastie, R.\ Tibshirani, and J.\ Friedman,
%   \textit{The Elements of Statistical Learning -- Data Mining, Inference, and Prediction}.
%   New York: Springer, 2009.
% \bibitem[4]{YourName17-XXX}
%   F.\ Lastname1, F.\ Lastname2, and F.\ Lastname3,
%   ``Title of your INTERSPEECH 2020 publication,''
%   in \textit{Interspeech 2020 -- 20\textsuperscript{th} Annual Conference of the International Speech Communication Association, September 15-19, Graz, Austria, Proceedings, Proceedings}, 2020, pp.~100--104.
% \end{thebibliography}

\end{document}